\documentclass{PoS}

\usepackage{latexsym}

\newcommand{\be}{\begin{equation}}
\newcommand{\ee}{\end{equation}}
\newcommand{\bea}{\begin{eqnarray}}
\newcommand{\eea}{\end{eqnarray}}
\newcommand{\mev}{\mbox{MeV}}
\newcommand{\gev}{\mbox{GeV}}
\newcommand{\VVP}{\langle V V \! P\rangle}
\def\order#1{{\cal O}\left(#1\right)}

\title{Hadronic light-by-light scattering in the muon $g-2$: a new
  short-distance constraint on pion exchange} 

\ShortTitle{Had.\ light-by-light scattering in $a_\mu$: a new
  short-distance constraint on $\pi^0$-exchange}  

\author{Andreas Nyffeler \\
        Regional Centre for Accelerator-based Particle Physics \\
        Harish-Chandra Research Institute \\
        Chhatnag Road, Jhusi \\
        Allahabad - 211 019, India \\ 
        E-mail: \email{nyffeler@hri.res.in}}

\abstract{We summarize our recent new evaluation of the pion-exchange
  contribution to hadronic light-by-light scattering in the muon $g-2$. We
  first derive a new short-distance constraint on the off-shell
  pion-photon-photon form factor at the external vertex in $a_\mu$ which
  relates the form factor to the quark condensate magnetic susceptibility in
  QCD. We then evaluate the pion-exchange contribution in the framework of
  large-$N_C$ QCD using an off-shell form factor which fulfills all
  short-distance constraints and obtain the new estimate
  $a_{\mu}^{\mathrm{LbyL};\pi^0} = (72 \pm 12) \times 10^{-11}$. Updating our
  earlier results for the contributions from the exchanges of the $\eta$ and
  $\eta^\prime$ using simple vector-meson dominance form factors, we get
  $a_{\mu}^{\mathrm{LbyL; PS}} = (99 \pm 16) \times 10^{-11}$ for the sum of
  all light pseudoscalars. Combined with available evaluations for the other
  contributions to hadronic light-by-light scattering this leads to the
  estimate $a_{\mu}^{\mathrm{LbyL; had}} = (116 \pm 40) \times 10^{-11}$. The
  corresponding contributions to the anomalous magnetic moment of the electron
  are also given.}

\PACS{13.40.Em, 12.38.Lg, 14.40.Aq, 14.60.Ef}

\FullConference{6th International Workshop on Chiral Dynamics\\
                July 6-10, 2009\\
                Bern, Switzerland}

\makeatletter
\setbox\@firstaubox\hbox{\small Andreas Nyffeler}
\makeatother

\begin{document}

\section{Introduction}
\label{sec:intro}

The muon $g-2$ has served over many decades as an important test of the
Standard Model (SM). It is also sensitive to contributions from New Physics
slightly above the electroweak scale. In fact, for several years now a
discrepancy of more than three standard deviations has existed between the SM
prediction and the experimental value, see the recent reviews
Refs.~\cite{MdeRR07, FJ_Reviews, JN09} on the muon $g-2$. The main error in
the theoretical SM prediction comes from hadronic contributions, i.e.\
hadronic vacuum polarization and hadronic light-by-light (had.\ LbyL)
scattering. Whereas the hadronic vacuum polarization contribution can be
related to the cross section $e^+ e^-\to~\mbox{hadrons}$, no direct
experimental information is available for had.\ LbyL scattering. One therefore
has to rely on hadronic models to describe the strongly interacting,
nonperturbative dynamics at the relevant scales from the muon mass up to about
2 GeV. This leads to large uncertainties, see Refs.~\cite{BP07, PdeRV09, JN09}
for recent reviews on had.\ LbyL scattering, largely based on the original
works~\cite{BPP95,HKS95_HK98,KNetal01,MV03}.

Essentially, these models describe the interactions of hadrons with photons,
usually with the help of some form factors.  One can reduce this model
dependence and the corresponding uncertainties by relating the form factors at
low energies to results from chiral perturbation theory (ChPT)~\cite{ChPT} and
at high energies (short distances) to the operator product expansion
(OPE)~\cite{OPE}. In this way, one connects the form factors to the underlying
theory of QCD. In particular, this has been done in Refs.~\cite{BPP95,
HKS95_HK98, Bijnens_Persson01, KNetal01, MV03} for the numerically dominant
contribution from the exchange of light pseudoscalars $\pi^0, \eta,
\eta^\prime$.

\begin{figure}[h]
\centerline{\includegraphics[width=.8\textwidth]{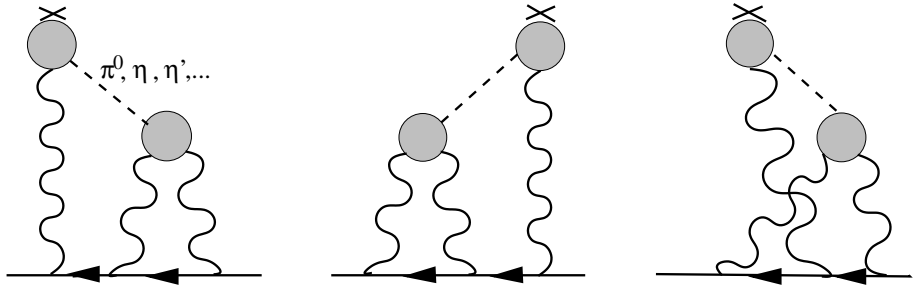}} 
\caption{The pseudoscalar-exchange contributions to had.\ LbyL 
  scattering. The shaded blobs represent the form factor ${\cal
  F}_{{\rm PS}^* \gamma^* \gamma^*}$ where ${\rm PS} = \pi^0, \eta,
  \eta^\prime, {\pi^0}^\prime, \ldots$.}
\label{fig:LbL_PS-exchanges}
\end{figure}

The pseudoscalar-exchange contributions to had.\ LbyL scattering are given by
the diagrams shown in Fig.~\ref{fig:LbL_PS-exchanges}.

It was pointed out recently in Ref.~\cite{FJ_Reviews}, that one should use
fully {\it off-shell} form factors for the evaluation of the LbyL scattering
contribution. This seems to have been overlooked in the recent literature, in
particular, in Refs.~\cite{Bijnens_Persson01,KNetal01,MV03,BP07,PdeRV09}. The
on-shell form factors as used in Refs.~\cite{KNetal01,Bijnens_Persson01}
actually violate four-momentum conservation at the external vertex, as
observed already in Ref.~\cite{MV03}.

The exchange of the lightest state $\pi^0$ yields the largest contribution and
therefore warrants special attention.  In these proceedings, based on the
results obtained in Ref.~\cite{Nyffeler09}, we present a new QCD
short-distance constraint on the off-shell pion-photon-photon form factor
${\cal F}_{{\pi^0}^* \gamma^* \gamma^*}$ at the external vertex by relating
it to the quark condensate magnetic susceptibility of QCD. We then evaluate
this contribution in the framework of large-$N_C$ QCD~\cite{largeNc}, using a
form factor which fulfills this new and other relevant short-distance
constraints.

\section{On-shell versus off-shell form factors} 
\label{sec:FF_onshell_offshell}

For the pion, the key object which enters the diagrams in
Fig.~\ref{fig:LbL_PS-exchanges} is the {\it off-shell} form factor ${\cal
F}_{{\pi^0}^*\gamma^*\gamma^*}((q_1+q_2)^2,q_1^2, q_2^2)$ which can be defined
via the QCD Green's function $\VVP$~\cite{BPP95,HKS95_HK98,Nyffeler09}
\bea
\lefteqn{ \int d^4 x\, d^4 y \, e^{i (q_1 \cdot x + q_2 \cdot y)} \, 
\langle\,0 | T \{ j_\mu(x) j_\nu(y) P^3(0) \} | 0 \rangle \, } \nonumber \\  
&=&  \varepsilon_{\mu\nu\alpha\beta} \, q_1^\alpha q_2^\beta \, 
 {i \langle{\overline\psi}\psi\rangle \over F_\pi} \, {i \over (q_1 +
   q_2)^2 - m_\pi^2} \, {\cal F}_{{\pi^0}^*\gamma^*\gamma^*}((q_1 +
 q_2)^2,q_1^2,q_2^2) \ + \ \ldots, 
\label{FFoffshellpi}
\eea
up to small mixing effects with the states $\eta$ and $\eta^\prime$ and
neglecting exchanges of heavier states like ${\pi^0}^\prime,
{\pi^0}^{\prime\prime}, \ldots$.  Here $j_\mu(x)$ is the light quark part of
the electromagnetic current and $P^3 = {\overline \psi} i \gamma_5 {\lambda^3
  \over 2} \psi$.  

The corresponding contribution to the muon $g-2$ may be worked out with the
result~\cite{KNetal01}  
\bea
\lefteqn{\hspace*{-1cm}a_{\mu}^{\mathrm{LbyL};\pi^0} =  - e^6
\int {d^4 q_1 \over (2\pi)^4} {d^4 q_2 \over (2\pi)^4}
\,\frac{1}{q_1^2 q_2^2 (q_1 + q_2)^2[(p+ q_1)^2 - m_\mu^2][(p - q_2)^2 -
    m_\mu^2]} 
\nonumber} \\
&& 
\times \left[
{{\cal F}_{{\pi^0}^*\gamma^*\gamma^*}(q_2^2, q_1^2, (q_1 + q_2)^2) \ {\cal 
    F}_{{\pi^0}^*\gamma^*\gamma}(q_2^2, q_2^2, 0) \over q_2^2 - 
m_{\pi}^2} \ T_1(q_1,q_2;p) \nonumber \right. \\
&& 
\quad  + \left. {{\cal F}_{{\pi^0}^*\gamma^*\gamma^*}((q_1+q_2)^2, q_1^2,
  q_2^2) \ {\cal F}_{{\pi^0}^*\gamma^*\gamma}((q_1+q_2)^2, (q_1+q_2)^2, 0)
  \over (q_1+q_2)^2 - m_{\pi}^2} \ T_2(q_1,q_2;p) \right], 
\label{a_pion_2}
\eea
where the external photon has now zero four-momentum. See Ref.~\cite{KNetal01}
for the expressions for $T_i$. 

Instead of the representation in Eq.~(\ref{a_pion_2}),
Refs.~\cite{Bijnens_Persson01,KNetal01} considered {\it on-shell} form factors
which would yield the so called {\it pion-pole} contribution, e.g.\ for the
term involving $T_2$, one would write~\cite{FJ_Reviews}
\be
{\cal F}_{\pi^0 \gamma^* \gamma^*}(m_\pi^2,
q_1^2,q_2^2) \ \times \ {\cal F}_{\pi^0 \gamma^* \gamma}(m_\pi^2,
(q_1 + q_2)^2,0). 
\ee
Although pole dominance might be expected to give a reasonable approximation,
it is not correct as it was used in those references, as stressed in
Refs.~\cite{MV03,FJ_Reviews}.  The point is that the form factor
sitting at the external photon vertex in the pole approximation ${\cal
F}_{\pi^0 \gamma^* \gamma}(m_\pi^2,(q_1 + q_2)^2,0)$ for $(q_1 + q_2)^2 \neq
m_\pi^2$ violates four-momentum conservation, since the momentum of the
external (soft) photon vanishes.  The latter requires ${\cal F}_{{\pi^0}^*
\gamma^* \gamma}((q_1 + q_2)^2, (q_1 + q_2)^2, 0)$. Ref.~\cite{MV03} then
proposed to use instead 
\be
{\cal F}_{\pi^0 \gamma^* \gamma^*}(m_\pi^2,
q_1^2,q_2^2) \ \times \ {\cal F}_{\pi^0 \gamma \gamma}(m_\pi^2,
m_\pi^2,0)\,. 
\ee
Note that putting the pion on-shell at the external vertex automatically leads
to a constant form factor, given by the Wess-Zumino-Witten (WZW)
term~\cite{WZW}.  However, this prescription does not yield the {\it
  pion-exchange} contribution with off-shell form factors, which we calculate
with Eq.~(\ref{a_pion_2}).    

Strictly speaking, the identification of the pion-exchange contribution is
only possible, if the pion is on-shell. If one is off the mass shell of the
exchanged particle, it is not possible to separate different contributions to
the $g-2$, unless one uses some particular model where elementary pions can
propagate. In this sense, only the pion-pole contribution with on-shell form
factors can be defined, at least in principle, in a model-independent way.  On
the other hand, the pion-pole contribution is only a part of the full result,
since in general, e.g.\ using some resonance Lagrangian, the form factors will
enter the calculation with off-shell momenta. In this respect, we view our
evaluation as being a part of a full calculation of had.\ LbyL scattering
using a resonance Lagrangian whose coefficients are tuned in such a way as to
systematically reproduce the relevant QCD short-distance constraints, along
the lines of the resonance chiral theory developed in Ref.~\cite{Ecker_etal}.

\section{A new short-distance constraint on the off-shell pion-photon-photon 
  form factor}
\label{sec:FF_constraints}

The form factor ${\cal
F}_{{\pi^0}^*\gamma^*\gamma^*}((q_1+q_2)^2,q_1^2,q_2^2)$ defined in
Eq.~(\ref{FFoffshellpi}) is determined by nonperturbative physics of QCD and
cannot (yet) be calculated from first principles. Therefore, various hadronic
models have been used in the literature. At low energies, the form factor is
normalized by the decay amplitude, ${\cal A}(\pi^0 \to \gamma\gamma) \equiv
e^2 {\cal F}_{\pi^0\gamma\gamma}(m_\pi^2, 0, 0)$. To a good approximation, all
hadronic models thus have to satisfy the constraint ${\cal
F}_{\pi^0\gamma\gamma}(m_\pi^2, 0, 0) = - N_C / (12 \pi^2 F_\pi)$.\footnote{We
note that in our work~\cite{Nyffeler09} and in Refs.~\cite{BPP95, HKS95_HK98,
KNetal01, MV03} simply $F_\pi = 92.4~\mev$ is used, without any error
attached. Maybe this could be an additional source of uncertainty in
$a_{\mu}^{\mathrm{LbyL};\pi^0}$, in particular in view of the new value
$\Gamma(\pi^0 \to \gamma\gamma) = (7.82 \pm 0.23)~\mbox{eV}$ presented in
Ref.~\cite{Bernstein09}; see also the discussion in
Ref.~\cite{Kampf_Moussallam09} and references therein.}

For an on-shell pion, there is also experimental data available for one
on-shell and one off-shell photon, from the process $e^+ e^- \to e^+ e^-
\pi^0$. Several experiments~\cite{CELLO90_CLEO98} thereby fairly well confirm
the Brodsky-Lepage~\cite{LepageBrodsky80} behavior for large Euclidean
momentum $\lim_{Q^2 \to \infty} \: {\cal F}_{\pi^0 \gamma^*
\gamma}(m_\pi^2,-Q^2,0) \sim - 2 F_\pi / Q^2$ and any model should reproduce
this behavior, maybe with a different prefactor.\footnote{Note, however, that
a recent measurement of the form factor by the BABAR
collaboration~\cite{BABAR09} at momentum transfers $Q^2$ between $4~\gev^2$
and $40~\gev^2$ does not show such a falloff. We will come back to this issue
in Section~\ref{sec:new_evaluation_pseudoscalars}.}

Apart from these experimental constraints, any consistent hadronic model for
the off-shell form factor ${\cal F}_{{\pi^0}^*\gamma^*\gamma^*}((q_1 + q_2)^2,
q_1^2, q_2^2)$ should match at large momentum with short-distance constraints
from QCD that can be calculated using the OPE. In Ref.~\cite{KN_EPJC01} the
short-distance properties for the three-point function $\VVP$ in
Eq.~(\ref{FFoffshellpi}) in the chiral limit and assuming octet symmetry have
been worked out in detail.  Two limits are of interest. In the first case, the
two momenta become simultaneously large, which describes the situation where
the space-time arguments of all three operators tend towards the same point at
the same rate. The second situation corresponds to the case where the relative
distance between only two of the three operators in $\VVP$ becomes small.
When the space-time arguments of the two vector currents in $\VVP$ approach
each other, the leading term in the OPE leads to the Green's function $\langle
A P\rangle$. The explicit results for both these cases can be found in
Refs.~\cite{KN_EPJC01,Nyffeler09}.

The new short-distance constraint on the off-shell form factor at the external
vertex in had.\ LbyL scattering arises when the space-time argument of one of
the vector currents in $\VVP$ approaches the argument of the pseudoscalar
density. This leads to the two-point function $\langle VT\rangle$ of the
vector current and the antisymmetric tensor density
\be \label{Pi_VT}
\delta^{ab}(\Pi_{\rm VT})_{\mu\rho\sigma}(p)\,=\, 
\int d^4x e^{i p \cdot x}
\langle 0 \vert T \{ V_\mu^a(x) 
({\overline\psi}\,\sigma_{\rho\sigma}\frac{\lambda^b}{2}\,\psi)(0)\}\vert
0\rangle \, , \qquad  \sigma_{\rho\sigma}={i\over
  2}[\gamma_{\rho},\gamma_{\sigma}]. 
\ee
Conservation of the vector current and invariance under parity then give
$(\Pi_{\rm VT})_{\mu\rho\sigma}(p)\,=
\,(p_{\rho}\eta_{\mu\sigma}-p_{\sigma}\eta_{\mu\rho})\,\Pi_{\rm VT}(p^2)$.  In
this way one obtains the relation (up to corrections of order
$\alpha_s$)~\cite{KN_EPJC01, Nyffeler09}
\be
\lim_{\lambda \to \infty} {\cal F}_{{\pi^0}^*\gamma^*\gamma^*}((\lambda q_1 +
q_2)^2, (\lambda q_1)^2,q_2^2) = - {2 \over 3} {F_0 \over
  \langle{\overline\psi}\psi\rangle_0} \Pi_{\rm VT}(q_2^2) 
+ \order{{1\over \lambda}} \, . \label{FF_OPE_3} 
\ee
In particular, at the external vertex in LbyL scattering in
Eq.~(\ref{a_pion_2}), the limit $q_2 \to 0$ is relevant. 

As pointed out in Ref.~\cite{Belyaev_Kogan}, the value of $\Pi_{\rm VT}(p^2)$
  at zero momentum is related to the quark condensate magnetic susceptibility
  $\chi$ in QCD in the presence of a constant external electromagnetic field,
  introduced in Ref.~\cite{Ioffe_Smilga}: $\langle 0 | \bar{q} \sigma_{\mu\nu}
  q | 0 \rangle_{F} = e \, e_q \, \chi \, \langle{\overline\psi}\psi\rangle_0
  \, F_{\mu\nu}$, with $e_u = 2/3$ and $e_d = -1/3$. With our definition of
  $\Pi_{\rm VT}$ in Eq.~(\ref{Pi_VT}) one obtains the relation $\Pi_{\rm
  VT}(0) = - (\langle{\overline\psi}\psi\rangle_0 / 2) \chi$ (see also
  Ref.~\cite{Mateu_Portoles}) and the new short-distance constraint at the
  external vertex can be written as~\cite{Nyffeler09}
\be
\lim_{\lambda \to \infty} {\cal F}_{{\pi^0}^*\gamma^*\gamma}((\lambda
q_1)^2, (\lambda q_1)^2,0) = {F_0 \over 3} \ \chi 
+ \order{{1\over \lambda}} .
\label{FF_OPE_3_zeromomentum_chi}  
\ee
Note that there is no falloff in this limit, unless $\chi$ vanishes.

Unfortunately there is no agreement in the literature what the actual value of
$\chi$ should be. In comparing different results one has to keep in mind that
$\chi$ actually depends on the renormalization scale $\mu$. In
Ref.~\cite{Ioffe_Smilga} the estimate $\chi(\mu = 0.5~\mbox{GeV}) = -
(8.16^{+2.95}_{-1.91})~\mbox{GeV}^{-2}$ was given in a QCD sum rule evaluation
of nucleon magnetic moments.  A similar value $\chi = - N_C / (4 \pi^2
F_\pi^2) = - 8.9~\mbox{GeV}^{-2}$ was obtained in Ref.~\cite{Vainshtein03},
probably again for a low scale $\mu \sim 0.5~\mbox{GeV}$ as argued in
Ref.~\cite{Vainshtein03}.

On the other hand, saturating the leading short-distance behavior of the
two-point function $\Pi_{\rm VT}~$\cite{Craigie:1981jx} with one multiplet of
lowest-lying vector mesons (LMD)~\cite{Balitsky_Yung,Belyaev_Kogan,KN_EPJC01}
leads to the estimate $\chi^{\rm LMD} = - 2 / M_V^2 =
-3.3~\mbox{GeV}^{-2}$~\cite{Balitsky_Yung}.  Again, it is not obvious at which
scale this relation holds, it might be at $\mu = M_V$.  This LMD estimate was
soon afterwards improved by taking into account higher resonance states
($\rho^\prime, \rho^{\prime\prime}$) in the framework of QCD sum rules, with
the results $\chi(0.5~\mbox{GeV}) = - (5.7 \pm
0.6)~\mbox{GeV}^{-2}$~\cite{Belyaev_Kogan} and $\chi(1~\mbox{GeV}) = - (4.4
\pm 0.4)~\mbox{GeV}^{-2}$~\cite{Balitsky_etal}. A more recent
analysis~\cite{Ball_etal} yields, however, a smaller absolute value
$\chi(1~\mbox{GeV}) = - (3.15 \pm 0.30)~\mbox{GeV}^{-2}$, close to the
original LMD estimate.\footnote{After the publication of our paper
Ref.~\cite{Nyffeler09}, two new estimates for $\chi$ appeared, both based on
the analysis of the zero-modes of the Dirac operator.  Ref.~\cite{Ioffe09}
presents an analytical approach which yields $\chi(1~\mbox{GeV}) = -
3.52~\mbox{GeV}^{-2}$ with an estimated error of $30-50\%$. A quenched lattice
calculation~\cite{Chi_Lattice} for $N_C = 2$ gives a very small absolute value
$\chi = - 1.547(6)~\mbox{GeV}^{-2}$. No scale dependence is given, the lattice
spacing corresponds to 2~GeV.}  For a quantitative comparison of all these
estimates for $\chi$ we would have to run them to a common scale, for
instance, 1 GeV or 2 GeV, which can obviously not be done within perturbation
theory starting from such low scales as $\mu = 0.5~\mbox{GeV}$.

\section{New evaluation of the pseudoscalar-exchange con\-tri\-bu\-tion in
  large-$N_C$ QCD}
\label{sec:new_evaluation_pseudoscalars}

In the spirit of the minimal hadronic Ansatz~\cite{MHA} for Green's functions
in large-$N_C$ QCD, an {\it off-shell} form factor ${\cal
F}_{{\pi^0}^*\gamma^*\gamma^*}((q_1+q_2)^2, q_1^2, q_2^2)$ has been
constructed in Ref.~\cite{KN_EPJC01}. It contains the two lightest multiplets
of vector resonances, the $\rho$ and the $\rho'$ (LMD+V), and fulfills all the
OPE constraints discussed earlier:
\bea
{\cal F}_{{\pi^0}^*\gamma^*\gamma^*}^{\rm LMD+V}(q_3^2, q_1^2, q_2^2)&=&
 \frac{F_\pi}{3}\, {q_1^2\,q_2^2\,(q_1^2 + q_2^2 + q_3^2) + P_H^V(q_1^2,
   q_2^2, q_3^2) \over (q_1^2-M_{V_1}^2) \, (q_1^2-M_{V_2}^2) \,
   (q_2^2-M_{V_1}^2) \, (q_2^2-M_{V_2}^2)} , 
\label{KNpipioff} \\  
P_H^V(q_1^2,q_2^2,q_3^2)&=& h_1\,(q_1^2+q_2^2)^2
+ h_2\,q_1^2\,q_2^2 + h_3\,(q_1^2+q_2^2)\,q_3^2 + h_4\,q_3^4 \nonumber \\  
&& +h_5\,(q_1^2+q_2^2) + h_6\,q_3^2 + h_7, 
\qquad \quad q_3^2 = (q_1 + q_2)^2. 
\eea

Below we reevaluate the pion-exchange contribution using off-shell LMD+V form
factors at both vertices. The constants $h_i$ in the Ansatz for ${\cal F}^{\rm
LMD+V}_{{\pi^0}^* \gamma^* \gamma^*}$ in Eq.~(\ref{KNpipioff}) are determined
as follows. The normalization with the pion decay amplitude $\pi^0 \to
\gamma\gamma$ yields $h_7 = - N_C M_{V_1}^4 M_{V_2}^4 / (4 \pi^2 F_\pi^2) -
h_6 m_\pi^2 - h_4 m_\pi^4 = -14.83~\mbox{GeV}^6 - h_6 m_\pi^2 - h_4 m_\pi^4$,
where we used $M_{V_1} = M_\rho = 775.49~\mbox{MeV}$ and $M_{V_2} =
M_{\rho^\prime} = 1.465~\mbox{GeV}$. The Brodsky-Lepage behavior can be
reproduced by choosing $h_1 = 0~\mbox{GeV}^2$. In Ref.~\cite{KN_EPJC01} a fit
to the CLEO data~\cite{CELLO90_CLEO98} for the on-shell form factor ${\cal
F}_{{\pi^0}\gamma^*\gamma}^{\rm LMD+V}(m_\pi^2, -Q^2, 0)$ was performed, with
the result $h_5 = (6.93 \pm 0.26)~\mbox{GeV}^4 - h_3 m_\pi^2$. The constant
$h_2$ can be obtained from higher-twist corrections in the OPE with the result
$h_2 = -10.63~\gev^2$~\cite{MV03}.

Within the LMD+V framework, the vector-tensor two-point function
reads~\cite{KN_EPJC01, Nyffeler09}  
\be
\Pi_{\rm VT}^{\rm LMD+V}(p^2) =   -\,\langle{\overline\psi}\psi\rangle_0\, 
{ p^2 + c_{\rm VT} \over (p^2-M_{V_1}^2) (p^2-M_{V_2}^2) }, \qquad \quad 
c_{\rm VT} = {M_{V_1}^2 M_{V_2}^2 \chi \over 2} . 
\ee
As shown in Ref.~\cite{KN_EPJC01}, the OPE constraint from Eq.~(\ref{FF_OPE_3})
  for ${\cal F}_{{\pi^0}^*\gamma^*\gamma^*}^{\rm LMD+V}$ leads to the relation 
\be \label{constraint_h1_h3_h4} 
h_1 + h_3 + h_4 = 2 c_{\rm VT} . 
\ee
The LMD estimate $\chi^{\rm LMD} = - 2 / M_V^2 = -3.3~\mbox{GeV}^{-2}$ is
close to $\chi(\mu = 1~\mbox{GeV}) = -(3.15 \pm 0.30)~\mbox{GeV}^{-2}$
obtained in Ref.~\cite{Ball_etal} using QCD sum rules with several vector
resonances $\rho, \rho^\prime$, and $\rho^{\prime\prime}$. Assuming that the
LMD/LMD+V framework is self-consistent, we will therefore take $\chi = (-3.3
\pm 1.1)~\mbox{GeV}^{-2}$ in our numerical evaluation, with a typical
large-$N_C$ uncertainty of about 30\%.  We will vary $h_3$ in the range $\pm
10~\mbox{GeV}^2$ and determine $h_4$ from Eq.~(\ref{constraint_h1_h3_h4}) and
vice versa.

The coefficient $h_6$ in the LMD+V Ansatz is undetermined as well. It enters
at order $p^6$ in the low-energy expansion of $\VVP$ in one combination of
low-energy constants from the chiral Lagrangian of odd intrinsic parity,
$A_{V,(p+q)^2}^{\rm LMD+V} = - F_\pi^2 h_6 / (8 M_{V_1}^4
M_{V_2}^4)$~\cite{KN_EPJC01}. The LMD ansatz with only one multiplet of vector
resonances yields $A_{V,(p+q)^2}^{\rm LMD} = - F_\pi^2 / (8 M_V^4) = -0.26 \
(10^{-4} / F_\pi^2)$~\cite{KN_EPJC01}. If the LMD/LMD+V framework is
self-consistent, the change in these estimates, while going from LMD to LMD+V,
should not be too big.  Since the size of this low-energy constant seems to be
small compared to another combination of low-energy constants which enters at
order $p^6$, we allow for a 100\% uncertainty of $A_{V,(p+q)^2}^{\rm LMD}$ and
get the range $h_6 = (5 \pm 5)~\mbox{GeV}^4$, see Ref.~\cite{Nyffeler09} for
details.

The results for $a_\mu^{\mathrm{LbyL};\pi^0}$ for some selected values of $h_3,
h_4$ and $h_6$, varied in the ranges discussed above, for $\chi =
-3.3~\mbox{GeV}^{-2}$, $h_1 = 0~\mbox{GeV}^2$, $h_2 = -10.63~\mbox{GeV}^2$ and
$h_5 = 6.93~\mbox{GeV}^4 - h_3 m_\pi^2$ are collected in
Table~\ref{tab:pi0res}, see Refs.~\cite{Nyffeler09, JN09} for details on the
numerics. 
\begin{table}[h] 
\begin{center} 
\begin{tabular}{|l|c|c|c|}
\hline 
 & ~$h_6 = 0~\mbox{GeV}^4$~ & ~$h_6 = 5~\mbox{GeV}^4$~ & ~$h_6 =
10~\mbox{GeV}^4$~ \\  
\hline 
~$h_3 = -10~\mbox{GeV}^2$~ & 68.4 & 74.1 & 80.2 \\ 
~$h_3 = 0~\mbox{GeV}^2$    & 66.4 & 71.9 & 77.8 \\
~$h_3 = 10~\mbox{GeV}^2$   & 64.4 & 69.7 & 75.4 \\ 
\hline 
~$h_4 = -10~\mbox{GeV}^2$  & 65.3 & 70.7 & 76.4 \\
~$h_4 = 0~\mbox{GeV}^2$    & 67.3 & 72.8 & 78.8 \\
~$h_4 = 10~\mbox{GeV}^2$   & 69.2 & 75.0 & 81.2 \\
\hline 
\end{tabular}
\end{center} 
\vspace*{-0.5cm} 
\caption{Results for $a_\mu^{\mathrm{LbyL};\pi^0}\times 10^{11}$ obtained with
the off-shell LMD+V form factor for $\chi = -3.3~\mbox{GeV}^{-2}$ and the
given values for $h_3, h_4$ and $h_6$. When varying $h_3$ (upper half of the
table), the parameter $h_4$ is fixed by the constraint in
Eq.~(\protect\ref{constraint_h1_h3_h4}). In the lower half the procedure is
reversed.} 
\label{tab:pi0res}
\end{table}

Varying $\chi$ by $\pm 1.1~\mbox{GeV}^{-2}$ changes the result for
$a_\mu^{\mathrm{LbyL};\pi^0}$ by $\pm 2.1 \times 10^{-11}$ at most. The
uncertainty in $h_6$ affects the result by up to $\pm 6.4 \times
10^{-11}$. The variation of $a_\mu^{\mathrm{LbyL};\pi^0}$ with $h_3$ [with
$h_4$ determined from the constraint in Eq.~(\ref{constraint_h1_h3_h4}) or
vice versa] is much smaller, at most $\pm 2.5 \times 10^{-11}$. In the absence
of more information on the values of the constants $h_3, h_4$ and $h_6$, we
take the average of the results obtained with $h_6 = 5~\mbox{GeV}^4$ for $h_3
= 0~\mbox{GeV}^2$ and for $h_4 =0~\mbox{GeV}^2$ as our central value: 
$a_\mu^{\mathrm{LbyL};\pi^0} = 72.3 \times 10^{-11}$.  Adding all
uncertainties from the variations of $\chi$, $h_3$ (or $h_4$), $h_5$ and $h_6$
linearly to cover the full range of values obtained with our scan of
parameters, we get~\cite{Nyffeler09, JN09}
\be \label{amupi0LMD+V}
a_\mu^{\mathrm{LbyL};\pi^0} = (72 \pm 12) \times 10^{-11}.
\ee
This value replaces the result obtained in Ref.~\cite{KNetal01} with on-shell
LMD+V form factors at both vertices.  We think the 16\% error should fairly
well describe the inherent model uncertainty using the {\it off-shell} LMD+V
form factor. In order to facilitate updates of our result in case some of the
parameters $h_i$ in the LMD+V Ansatz in Eq.~(\ref{KNpipioff}) will be known
more precisely, we have given in the Appendix of Ref.~\cite{Nyffeler09} a
parametrization of $a_\mu^{\mathrm{LbyL};\pi^0}$ for arbitrary coefficients
$h_i$.\footnote{A fit of the on-shell LMD+V form factor to the recent BABAR
data~\cite{BABAR09} yields $h_1 = (-0.17 \pm 0.02)~\mbox{GeV}^2$ and $h_5 =
(6.51 \pm 0.20)~\mbox{GeV}^4 - h_3 m_\pi^2$ with $\chi^2/\mbox{dof} = 15.0 /
15 = 1.0$. In this way we would get the new average value
$a_\mu^{\mathrm{LbyL};\pi^0} = 71.8 \times 10^{-11}$, i.e.\ the result is
essentially unchanged from Eq.~(\ref{amupi0LMD+V}).}

As far as the contribution to $a_\mu$ from the exchanges of the other light
pseudoscalars $\eta$ and $\eta^\prime$ is concerned, it is not so
straightforward to apply the above analysis within the LMD+V framework to
these resonances. In particular, the short-distance analysis in
Ref.~\cite{KN_EPJC01} was performed in the chiral limit and assumed octet
symmetry. We therefore resort to a simplified approach which was also adopted
in other works~\cite{BPP95,HKS95_HK98,KNetal01,MV03} and take a simple VMD
form factor normalized to the experimental decay width $\Gamma(\mbox{PS} \to
\gamma \gamma)$. In this way we obtain the results
$a_{\mu}^{\mathrm{LbyL};\eta} = 14.5 \times 10^{-11}$ and
$a_{\mu}^{\mathrm{LbyL};\eta^\prime} = 12.5 \times 10^{-11}$, which update the
values given in Ref.~\cite{KNetal01}. Adding up the contributions from all the
light pseudoscalar exchanges, we obtain the estimate~\cite{Nyffeler09, JN09} 
\be \label{amuLbLPS}
a_{\mu}^{\mathrm{LbyL;PS}} = (99 \pm 16) \times 10^{-11}, 
\ee
where we have assumed a 16\% error, as inferred above for the pion-exchange
contribution.\footnote{Applying the same procedure to the electron, we get
$a_e^{\mathrm{LbyL};\pi^0} = (2.98 \pm 0.34) \times
10^{-14}$~\cite{Nyffeler09}. This number supersedes the value given in
Ref.~\cite{KNetal01}. Note that the naive rescaling
$a_e^{\mathrm{LbyL};\pi^0}(\mathrm{rescaled}) = (m_e / m_\mu)^2 \
a_\mu^{\mathrm{LbyL};\pi^0} = 1.7 \times 10^{-14}$ yields a value which is
almost a factor of 2 too small. Our estimates for the other pseudoscalars
contributions using VMD form factors at both vertices are
$a_e^{\mathrm{LbyL};\eta} = 0.49 \times 10^{-14}$ and
$a_e^{\mathrm{LbyL};\eta^\prime} = 0.39 \times 10^{-14}$. Therefore we get
$a_e^{\mathrm{LbyL;PS}} = (3.9 \pm 0.5) \times 10^{-14}$, where the relative
error of about 12\% is again taken over from the pion-exchange
contribution. Assuming that the pseudoscalar contribution yields the bulk of
the result of the total had.\ LbyL scattering correction, we obtain
$a_e^{\mathrm{LbyL;had}} = (3.9 \pm 1.3) \times 10^{-14}$, with a conservative
error of about 30\%, see Ref.~\cite{JN09}. This value was later confirmed in
the published version of Ref.~\cite{PdeRV09} where a leading logs estimate
yielded $a_e^{\mathrm{LbyL;had}} = (3.5 \pm 1.0) \times 10^{-14}$.}

\section{Discussion and conclusions}
\label{sec:conclusions} 
 
We would like to stress that although our result for the pion-exchange
contribution is not too far from the value
$a_\mu^{\mathrm{LbyL};\pi^{0}-\mathrm{pole}} = (76.5 \pm 6.7) \times 10^{-11}$
given in Ref.~\cite{MV03}, this is {\it pure coincidence}. We have used
off-shell LMD+V form factors at both vertices, whereas the authors of
Ref.~\cite{MV03} evaluated the {\it pion-pole} contribution using the on-shell
LMD+V form factor ${\cal F}_{{\pi^0}\gamma^*\gamma^*}(q_1^2, q_2^2) \equiv
{\cal F}_{{\pi^0}\gamma^*\gamma^*}(m_\pi^2, q_1^2, q_2^2)$ at the internal
vertex and a constant WZW form factor at the external vertex, see for instance
Eq.~(18) in Ref.~\cite{MV03}.  Since only the pion-pole contribution is
considered in Ref.~\cite{MV03}, their short-distance constraint cannot be
applied to our approach either. However, our ansatz for the pion-exchange
contribution agrees qualitatively with the short-distance behavior of the
quark-loop derived in Ref.~\cite{MV03}, see the discussion in
Refs.~\cite{Nyffeler09, JN09}. 

Our results for the pion and the sum of all pseudoscalars are about 20\%
larger than the values in Refs.~\cite{BPP95, HKS95_HK98} which used other
hadronic models. An evaluation of the pion-exchange contribution using
an off-shell form factor based on a nonlocal chiral quark model yielded
$a_{\mu}^{\mathrm{LbyL};\pi^0} = (65 \pm 2) \times
10^{-11}$~\cite{Dorokhov_Broniowski}. In that model, off-shell effects of the
pion always lead to a strong damping in the form factor and the result is
therefore smaller than the pion-pole contribution obtained in
Ref.~\cite{MV03}. In our model, there are some corners of the parameter space
where the result is larger than the pion-pole contribution, for instance, we
get a maximal value of $a_\mu^{\mathrm{LbyL};\pi^0} = 83.3 \times 10^{-11}$ in
the scanned region.  Very recently, a value of $a_{\mu}^{\mathrm{LbyL;PS}} =
107 \times 10^{-11}$ with an estimated error of at most 30\% was obtained in
Ref.~\cite{AdSQCD} within an AdS/QCD approach.

Combining our result for the pseudoscalars with the evaluation of the
axial-vector contribution in Ref.~\cite{MV03} and the results from
Ref.~\cite{BPP95} for the other contributions, we obtain the
estimate~\cite{Nyffeler09, JN09}  
\be
a_{\mu}^{\mathrm{LbyL; had}} = (116 \pm 40) \times 10^{-11} 
\ee
for the total had.\ LbyL scattering contribution to the anomalous magnetic
moment of the muon. To be conservative, we have added all the errors linearly,
as has become customary in recent years. In the very recent
review~\cite{PdeRV09} the central values of some of the individual
contributions to had.\ LbyL scattering are adjusted and some errors are
enlarged to cover the results obtained by various groups which used different
models. The errors are finally added in quadrature to yield the estimate
$a_{\mu}^{\mathrm{LbyL; had}} = (105 \pm 26) \times 10^{-11}$. Note that the
dressed light quark loops are not included as a separate contribution in
Ref.~\cite{PdeRV09}. They are assumed to be already covered by using the
short-distance constraint from Ref.~\cite{MV03} on the pseudoscalar-pole
contribution. Certainly, more work on the had.\ LbyL scattering contribution
is needed to fully control all the uncertainties.

\acknowledgments

I would like to thank the organizers of Chiral Dynamics 2009 for their
financial support and for providing such a stimulating atmosphere.  I am
grateful to F.\ Jegerlehner for pointing out that fully off-shell form factors
should be used to evaluate the pion-exchange contribution, for helpful
discussions and for numerous correspondences. Furthermore, I would like to
thank G.\ Colangelo, J.\ Gasser, M.\ Knecht, H.\ Leutwyler, P.\ Minkowski, B.\
Moussallam, M.\ Perrottet, A.\ Pich, J.\ Portoles, J.\ Prades, E.\ de Rafael
and A.\ Vainshtein for illuminating discussions. This work was supported by
the Department of Atomic Energy, Government of India, under a 5-Years Plan
Project.

\end{document}